# Magnetotransport Coefficients of $Sm_{0.55}Sr_{0.45}MnO_3$

R. Suryanarayanan[a] and V. Gasumyants[b]

[b]*Laboratoire de Physico-Chimie de l'Etat Solide, CNRS, UMR 8648, Bât 414 Université Paris-Sud, 91405 Orsay, France*

[b]*Dept. Of Semiconductor Physics, St.Petersburg Sytate Technical University, 29 Polytechnicheskaya, 195251 St.Petersburg, Russia*

Abstract

Measurements of Seebeck effect in 0 and 1.5 T and Nernst coefficient in 0.3, 0.9 and 1.8 T as a function of temperature on a polycrystalline sample $Sm_{0..55}Sr_{0.45}MnO_3$ are presented. The data point out conclusively that an increase in both the carrier density and the mobility of the charge carriers is responsible for the observed colossal magnetoresistance in this compound.

It is important to investigate the various magneto-transport coefficients such as Hall($R_H$) Seebeck(S) and Nernst(Q) of manganites in order to understand some of the mechanisms underlying the colossal magnetorsistance (CMR) effect in these compounds. Whereas, a large number of data exist on $R_H$ and S of these compounds[1], very little has been reported [2] on measurements of Q. Briefly, Nernst coefficient is the ratio of the transverse electric field component to the thermal gradient and magnetic field and is independent of the sign of the carriers. The simplest theory in the case of metallic conduction with a single type of carrier, spherical Fermi surface and isotropic relaxation time is given by [3]

$$Q = (\pi^2 k^2 T/3m) (d\tau(\varepsilon)/d\varepsilon)_{\varepsilon = \varepsilon_F},$$

where k is the Boltzmann constant, m is the effective mass and $\tau(\varepsilon)$ is the energy dependent relaxation time of the carrier near the Fermi energy $\varepsilon_F$. However, a lack of knowledge of the detailed energy dependence of the relaxation time, among other factors,. makes the quantitative analysis of the Nernst coefficient difficult.

We report here, magnetization(M), resistivity ($\rho$), S and Q measurements as a function of temperature(T) and magnetic field(B), of a polycrystalline sample $Sm_{0.55}Sr_{0.45}MnO_3$.(a=5.442,b=5.441,c=7.672 Â, Pbnm)

Here, we focus on our data and the discussion only near $T_c$, where the CMR effect is large. The experimental details are described earlier [2]. The $Sm_{0.55}Sr_{0.45}MnO_3$. sample shows a paramagnetic to ferromagnetic transition with a $T_c$ ~140 K accompanied by a semiconductor to metal transition (fig.1 and inset). $\rho$ shows a maximum at $T_c$ and decreases in B=1.8 T, indicating a sizeable CMR effect. Fig. 2 shows S as a function of T. In B=0, the value of S sharply increases as T approaches $T_c$ and for T<130 K, it increases rather slowly. In B=1.5 T, S starts increasing at T(=160 K) >$T_c$ and shows a small increase from T=150K. The increase in S mainly reflects an increase in the number of charge carriers which can account partly for the CMR. However, the mobility can also increase. Fig. 3 shows Q as a function of T in B=0.3, 0.9 and 1.8 T. For 160<T<300 K, Q is independent of B. In the cse of small polarons, the mobility, u ~(1/T) exp (-W / kT). Assuming Q $\propto \tau$, the scattering time and if u $\propto \tau$, then the absolute value of Q should increase when T decreases [2]. However, the value of Q is remarkably suppressed as B increased from 0.3 to 1.8 T for 100<T<150 K. This suppression in Q can be attributed to a changeover from a small polaron behavior to a large polaron behavior with a concomitant reduction in $\tau$, and hence to an increase in the mobility of the delocalized carriers. To explicitly illustrate this, we plot in

fig.4, CMR $=-100 \times (\rho_{1.8} - \rho_0) / \rho_0$, CSE$=100 \times (S_{1.5} - S_0) / S_0$ and CNE$=-100 \times (Q_{1.8} - Q_{0.3}) / Q_{0.3}$, as a function of T. The following features can be noticed. First, the value of CMR increases rapidly for T>160 K reaches a maximum of 75% at 140 K and then decreases rather slowly. Next, the value of CSE also increases for T>160 K, reaches a maximum earlier than the CMR, at 150 K, stays constant till 140 K and then reduces rapidly. Now, if we look at CNE, it increases only for T>155 K and reaches a maximum at T near $T_c$ where the CMR also has a maximum value. For T>140 K, the value of CNE does not change. These observations imply that the CMR effect (i) for 160 <T<140 K, is mostly due to a a rapid increase in the number of charge carriers and (ii) for 140<T<150 K, is due to an increase in the mobility of charge carriers. The increase in the mobility arises out of reduced spin-scattering in the presence of a magnetic field. We briefly recall that a mechanism was proposed [4] based on an exchange interaction of polaronic carriers with localized spins according to which it is the carrier collapse that accounts for the resistivity peak and the CMR effect. Our data conclusively demonstrate that the increase in mobility plays a key role and hence the proposed model [4] should take this aspect into account.

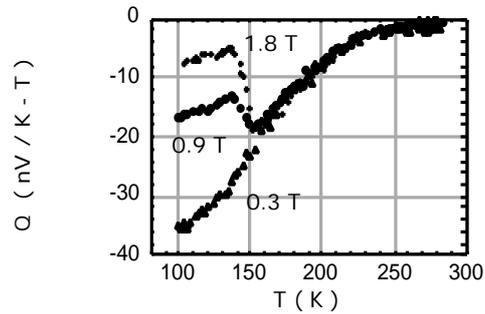

Figure 3. Q vs T in B= 0.3, 0.9 and 1.8 T.

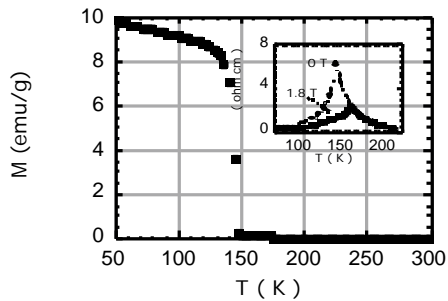

Figure 1. M vs T in B=0.01 T. (inset) $\rho$ vs T in B=0 and 1.8 T.

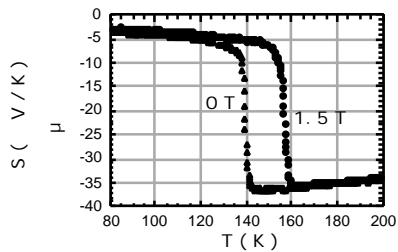

Figure 2. S vs T in B= 0 and 1.5 T.

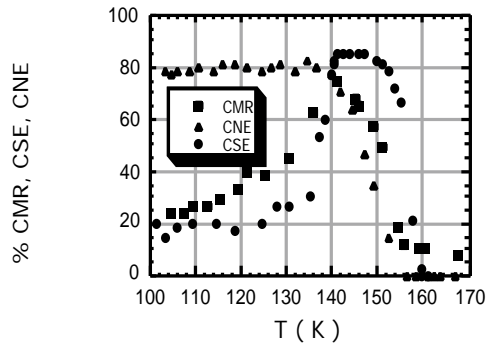

Figure 4. CMR $=-100 \times (\rho_{1.8} - \rho_0) / \rho_0$, CSE$=100 \times (S_{1.5} - S_0) / S_0$ and CNE$=-100 \times (Q_{1.8} - Q_{0.3}) / Q_{0.3}$, as a function of T

Acknowledgement

VG would like to thank the Russian Foundation for Basic Resaerch (Grant No. 02-02-1684).

References
[1] M. B. Salamon and M. Jaime, Rev. Mod. Phys. 73 (2001) 583 and references therein.
[2] R. Suryanarayanan, V. Gasumyants and N. Ageev, Phys. Rev. B 59 (1999) R9019.
[3] F. J. Blatt Solid State Phys. 4 (1957) 200
[4] A. S. Alexandrov and A. M. Bratkovsky, Phys. Rev. Lett. 82 (1999) 141.